\documentclass[aps,prd,twocolumn,nofootinbib,superscriptaddress]{revtex4-2}

\usepackage{amsmath,amssymb,amsfonts}
\usepackage{graphicx}
\usepackage{bm}
\usepackage{booktabs}
\usepackage{hyperref}
\usepackage{xcolor}

\begin{document}

\title{Constraining on Buchdahl-Inspired Gravity from Future Pulsar Timing near Sgr A*}

\author{Jian-Ming Yan}
\email{yanjm@ucas.ac.cn}
\affiliation{School of Fundamental Physics and Mathematical Sciences, Hangzhou Institute for Advanced Study, University of Chinese Academy of Sciences, Hangzhou 310024, China}

\author{Tao Zhu$^*$}
\email{zhut05@zjut.edu.cn}
\affiliation{Institute for Theoretical Physics \& Cosmology, Zhejiang University of Technology, Hangzhou, 310023, China}
\affiliation{United Center for Gravitational Wave Physics (UCGWP),  Zhejiang University of Technology, Hangzhou, 310023, China}

\author{Zong-Kuan Guo}
\email{guozk@itp.ac.cn}
\affiliation{School of Fundamental Physics and Mathematical Sciences, Hangzhou Institute for Advanced Study, University of Chinese    Academy of Sciences, Hangzhou 310024, China}
\affiliation{Institute of Theoretical Physics, Chinese Academy of Sciences (CAS), Beijing 100190, China}
\affiliation{University of Chinese Academy of Sciences (UCAS), Beijing 100049, China}

\author{Zhao Li}
\email{lz111301@mail.ustc.edu.cn}
\affiliation{Department of Astronomy, Peking University, Beijing, 100871, China}

\begin{abstract}
Future pulsar timing observations near Sgr~A* offer a unique probe of gravitational physics in the vicinity of a supermassive black hole.
We forecast the ability of such measurements to constrain a Buchdahl-inspired $R^2$ gravity, parameterized by a single deviation parameter $\epsilon$, using a timing framework that self-consistently integrates orbital dynamics with light-propagation delays and preserves the full timing solution across the observing span.
Through Fisher-matrix forecasts for a representative pulsar, we systematically isolate how the precision on $\epsilon$ depends on orbital geometry.
We find that shorter orbital periods and higher eccentricities significantly enhance sensitivity, consistent with a substantial contribution from observations near periastron.
As a benchmark comparison, we further consider a hypothetical pulsar on an S2-like orbit ($P_b=16~{\rm yr}$, $e=0.88$) and obtain a statistical sensitivity of $\sigma_\epsilon\sim 10^{-4}$ within the adopted weak-field, static, and spherically symmetric timing model.
This sensitivity is comparable to the natural order-of-magnitude truncation scale of the 1PN expansion and should not be interpreted as a complete forecast for the real Sgr~A* system.
Under the adopted idealized assumptions, the characteristic statistical scale is several orders of magnitude below the current S2 95\% confidence interval half-width ($|\epsilon|_{\rm S2}^{\rm 95\%}\approx 0.56$), though this comparison is heuristic given the differing confidence levels.
These trends provide quantitative guidance for target selection and campaign design in future Galactic-center pulsar searches.
\end{abstract}

\maketitle

\section{Introduction}
\label{sec:introduction}

General Relativity (GR) provides a remarkably successful description of gravitational phenomena and has passed a wide range of tests in both the solar system and compact-object environments~\cite{Will2014}. It predicts compact-object observables such as black-hole shadows and the dynamics of binary systems probed by gravitational waves~\cite{eht2019, eht2022, ligo2016-1, ligo2016-2}. Nevertheless, GR also motivates open questions, including its relationship to quantum physics and the consistency of gravitational theory across scales. These motivations have led to a variety of modified-gravity models, which must be confronted with increasingly precise observations.

Among modified theories, $f(R)$ gravity constitutes one of the most extensively studied extensions of Einstein's theory, encompassing a broad class of higher-curvature models whose phenomenology ranges from cosmological acceleration to compact-object spacetimes~\cite{SotiriouFaraoni2010, DeFeliceTsujikawa2010, Nojiri2011, Nojiri2017, Astashenok2013, Astashenok2015, Astashenok2017}. A particularly important subclass is pure $R^2$ gravity, in which the gravitational action is proportional to the square of the Ricci scalar. Quadratic-curvature gravity more broadly has been shown to be power-counting renormalizable in the foundational work of Stelle~\cite{Stelle1977, Stelle1978}, and pure $R^2$ gravity can also be embedded into supergravity~\cite{Gaume2016}. However, obtaining and interpreting asymptotically flat solutions is nontrivial because the field equations are of fourth order in the metric.

Buchdahl studied static, spherically symmetric vacuum solutions in $R^2$ gravity~\cite{Buchdahl1962}. For asymptotically flat solutions, the resulting spacetime includes a Buchdahl parameter $k$ and reduces to Schwarzschild when $k=0$. Recently, Nguyen constructed the relevant vacuum solutions in a compact and exhaustive form~\cite{Nguyen2022, Nguyen2023}. The phenomenological implication is that the Buchdahl-inspired metric provides a controlled deformation away from Schwarzschild while retaining a well-defined GR limit.

Naturally, this new solution requires phenomenological validation against observations. The Galactic center, hosting the supermassive black hole (SMBH) Sgr~A* at a distance of approximately $8~{\rm kpc}$, provides an ideal laboratory for gravity tests in the strong-gravity regime. The S-star cluster, and in particular the bright star S2 with its 16-year orbital period and high eccentricity ($e\approx 0.88$), has enabled milestone relativistic measurements. The GRAVITY instrument on the Very Large Telescope Interferometer has detected both the gravitational redshift~\cite{GRAVITY2018} and the Schwarzschild precession~\cite{GRAVITY2} in the S2 orbit, establishing the Galactic center as a precision probe of relativistic gravity.

More recently, the GRAVITY collaboration has discovered the star S301, which orbits Sgr~A* with a period of only $8.7~{\rm yr}$ and reaches a peak velocity of $\sim 25{,}000~{\rm km/s}$ at a pericenter distance of merely $\sim 12~{\rm AU}$~\cite{S301}. This makes S301 the closest known star to the black hole and opens the prospect of measuring the spin of Sgr~A* through the Lense--Thirring frame-dragging effect within the next decade~\cite{S301,S301spin}. The discovery of S301 further underscores the Galactic center as a rapidly advancing laboratory for relativistic gravity, and illustrates the continuing potential of stellar-orbit measurements to probe spacetime geometry near Sgr~A*.

In our previous work, we tested the Buchdahl-inspired $R^2$ metric against solar-system experiments, the S2 orbit, and the shadow of Sgr~A*, obtaining constraints on the deviation parameter~\cite{Zhu2024, Yan2024, Yan2025}. However, due to current limitations in astrometric and spectroscopic precision, these constraints are not yet competitive. For example, the S2 analysis yields a 95\% confidence interval half-width of $|\epsilon|\lesssim 0.56$~\cite{Yan2024}. This motivates the search for more sensitive observational tracers near Sgr~A*.

Pulsar timing offers such a tracer. Since the discovery of the Hulse--Taylor binary pulsar PSR~B1913+16~\cite{HulseTaylor1975, Taylor:1994zz}, radio pulsars have served as precision clocks for testing relativistic gravity, culminating in compact-binary tests with the Double Pulsar system~\cite{Kramer2021}. The Galactic center is expected to host a large population of pulsars; estimates based on multiwavelength constraints suggest that of order $10^3$ pulsars and potentially $\sim 10^2$ millisecond pulsars may reside within the central degree of the Galaxy~\cite{Wharton2012}. The SKA Observatory and the Next Generation Very Large Array (ngVLA) are expected to discover pulsars in the Galactic center, potentially including relativistic systems in close orbit around Sgr~A*~\cite{Kramer2004, Bower:2018mta}. Compared with stellar-orbit measurements such as those of S2~\cite{deMartino:2021daj,DellaMonica:2021xcf,DellaMonica:2021fdr,DAddio:2021smm,nohair,nohair2,Benisty:2021cmq,Han:2014yga,Yan:2023vdg,Yan:2022fkr,Zhang:2024fpm,DeMartino:2023qkl,DellaMonica:2023dcw,deLaurentis:2022oqa,Fernandez:2023kro,Cadoni:2022vsn,Shaymatov:2023jfa,GRAVITY2,GRAVITY2018,Mon}, pulsar timing can attain substantially higher effective precision through phase-coherent tracking of pulse signals. In particular, the improved high-frequency sensitivity of these facilities is crucial because strong interstellar scattering toward the Galactic center suppresses detectability at lower radio frequencies~\cite{Bower:2018mta}.

A growing literature has investigated how Galactic-center pulsar observations can strengthen tests of gravity near Sgr~A* and constrain departures from GR in the vicinity of the black hole~\cite{Liu:2011ae, Bower:2018mta, Yu:2025apk, Shao:2025vmb, Hu:2024blq, Hu:2023ubk, DellaMonica:2023ydm, Dong:2022zvh, Hu:2026zcb}. Psaltis, Wex, and Kramer outlined a program to quantitatively test the no-hair theorem by combining stellar orbits, pulsar timing, and black-hole shadow imaging~\cite{nohair}. The spin of Sgr~A* can in principle be measured via relativistic frame dragging imprinted on pulsar time of arrival (TOAs), and Hu and Shao have shown that two pulsars suffice to break key degeneracies in such a measurement~\cite{Hu:2024blq}. More broadly, pulsar timing in the Galactic center can probe alternative theories of gravity, dark matter, and spacetime geometry near a supermassive black hole~\cite{DellaMonica:2023ydm, Dong:2022zvh, Yu:2025apk}.

This motivates our Fisher-matrix forecasts for how accurately future pulsar observations in the Galactic center can constrain deviations from the Schwarzschild benchmark in Buchdahl-inspired $R^2$ gravity. In this work, we focus on the ability of future pulsar observations to constrain the deviation parameter $\epsilon$, introduced through a Buchdahl-inspired modification of the metric. Rather than analyzing a single assumed system, we ask a selection question: which orbital configurations maximize the information content for $\epsilon$ among prospective Galactic-center pulsars?

To answer this question, we carry out Fisher-matrix forecasts under controlled observational assumptions. By varying orbital parameters, we isolate how the geometry-dependent timing signatures translate into changes in the uncertainty on $\epsilon$. Additionally, to enable a direct comparison between pulsar timing and stellar-orbit methods, we forecast the precision achievable for a hypothetical pulsar on an S2-like orbit and assess the improvement relative to current S2-based constraints.

The paper is organized as follows. In Sec.~\ref{sec:theory} we summarize the Buchdahl-inspired $R^2$ metric and specify how $\epsilon$ enters the orbital dynamics and propagation delays. In Sec.~\ref{sec:fisher} we introduce the Fisher-matrix formalism used to forecast uncertainties and correlations. The fiducial numerical setup is given in Sec.~\ref{sec:numerical_setup}, and the resulting trends for $\epsilon$ as functions of orbital parameters and observing time are presented in Sec.~\ref{sec:results}, including a discussion of the domain of validity of the weak-field forecast in Sec.~\ref{sec:model_validity}. We conclude in Sec.~\ref{sec:conclusion}. Throughout, we use geometrized units with $G_{\rm N}=c=1$, unless otherwise stated, so that $M_\bullet$ denotes the gravitational mass measured by a distant observer.

\section{Theoretical Framework}
\label{sec:theory}

\subsection{Buchdahl-inspired $R^2$ correction}
\label{sec:buchdahl_metric}

We consider an asymptotically flat Buchdahl-inspired vacuum solution in pure $R^2$ gravity, which provides a non-Schwarzschild extension of the Schwarzschild spacetime~\cite{Nguyen2022, Nguyen2023}. Besides the central mass, the solution contains an additional integration constant, the Buchdahl parameter $k$. The Schwarzschild solution, and hence the GR limit, is recovered for $k=0$.

Following Ref.~\cite{Yan2024}, the deviation from Schwarzschild spacetime is described by the dimensionless parameter
\begin{equation}
    \epsilon \equiv \frac{\tilde{k}+\eta}{1+\eta},
    \label{eq:epsilon_definition}
\end{equation}
where $\tilde{k}=k/r_s$ is the dimensionless Buchdahl parameter, $r_s$ is the Schwarzschild radius, and $\eta$ is a second free parameter entering the metric in the weak-field expansion. In the general weak-field family, $\eta$ and $\tilde{k}$ are independent; however, the requirement that the metric match the asymptotically flat Buchdahl solution of Refs.~\cite{Nguyen2022, Nguyen2023} fixes $\eta=\tilde{k}$. With this identification,
\begin{equation}
    \epsilon = \frac{2\tilde{k}}{1+\tilde{k}},
    \label{eq:epsilon_ktilde}
\end{equation}
so that $\epsilon=0$ if and only if $\tilde{k}=0$ (i.e., $k=0$), confirming that the GR/Schwarzschild limit corresponds to $\epsilon=0$. In the Newtonian limit, the parameter $\eta$ also enters through a renormalization of the gravitational constant, $(1+\eta)G = G_{\rm N}$, where $G_{\rm N}$ is the Newtonian constant measured by a distant observer~\cite{Yan2024}.

A key subtlety concerns the physical interpretation of the mass parameter. Distant orbital observations (whether stellar or pulsar) measure the Newtonian gravitational coupling $\mu_{\rm N} \equiv G_{\rm N} M_{\rm physical}$, i.e., the product of the Newtonian constant and the physical mass of the central object. Since $\eta$ renormalizes $G$ into $G_{\rm N}$, the combination $\mu_{\rm N} = (1+\eta)G\,M$ absorbs the $\eta$-dependence: variations in $\eta$ (and hence in $\epsilon$) do not affect the Newtonian coupling $\mu_{\rm N}$. In what follows, we therefore adopt $\mu_{\rm N}$ as the mass parameter entering the metric, Kepler's law, and the equations of motion. With this convention, $\epsilon$ enters exclusively through the 1PN and light-propagation terms, while the Newtonian dynamics depend only on $\mu_{\rm N}$ and are independent of $\epsilon$.

In the following, $\epsilon$ is treated as a phenomenological modified-gravity parameter to be constrained by pulsar timing observations, with $\epsilon=0$ marking the GR limit. In geometrized units ($G_{\rm N}=c=1$), the mass parameter $\mu_{\rm N}$ is denoted $M_\bullet$ throughout.

In the weak-field region relevant to pulsars orbiting Sgr~A*, the metric can be expressed in isotropic coordinates $(t,r,\theta,\phi)$ as
\begin{equation}
    ds^2 =
    -f(r)\,dt^2
    +g(r)\left[
        dr^2+r^2\left(d\theta^2+\sin^2\theta\,d\phi^2\right)
    \right],
    \label{eq:weakfield_metric}
\end{equation}
where, retaining terms through the first post-Newtonian order,
\begin{align}
    f(r) &=
    1-\frac{2M_\bullet}{r}
    +\frac{2M_\bullet^2}{r^2},
    \label{eq:metric_f}\\
    g(r) &=
    1+2(1-\epsilon)\frac{M_\bullet}{r}.
    \label{eq:metric_g}
\end{align}
For $\epsilon=0$, these expressions reduce to the weak-field expansion of the Schwarzschild metric in isotropic coordinates.

At the order considered here, the Buchdahl-inspired correction appears in the spatial part of the metric, whereas the temporal component retains its Schwarzschild form.
The Shapiro delay coefficient can be obtained directly from the null geodesics of the metric~\eqref{eq:weakfield_metric}, without invoking the PPN formalism.
Setting $ds^2=0$ for a radially propagating photon and working through first post-Newtonian order, the coordinate speed of light acquires a factor $\sqrt{g/f}$, whose logarithmic integral over the path yields a delay proportional to the spatial-curvature coefficient $(1-\epsilon)$ in $g(r)$ together with the temporal potential coefficient in $f(r)$.
The two contributions combine to give a total Shapiro coefficient
\begin{equation}
    1+(1-\epsilon)=2-\epsilon,
    \label{eq:shapiro_coefficient}
\end{equation}
where the first term ($1$) arises from the temporal metric component $f(r)$ and the second ($1-\epsilon$) from the spatial metric component $g(r)$.
For reference, this corresponds to $\gamma_{\rm PPN}=1-\epsilon$ and $\beta_{\rm PPN}=1$ in the standard PPN notation~\cite{Will2014}. 

However, the PPN framework is not directly applicable to the Buchdahl-inspired metric. The correspondence $\gamma_{\rm PPN}=1-\epsilon$ noted above invites the question: does the solar-system bound $|\gamma_{\rm PPN}-1|\lesssim 10^{-5}$~\cite{Will2014} already constrain $\epsilon$ more tightly than the pulsar-timing forecast?
Solar-system bounds cannot be straightforwardly translated into bounds on the object-dependent Sgr~A* value of $\epsilon$, for two reasons.

First, $\epsilon$ is not a universal PPN parameter.
It is defined through Eq.~\eqref{eq:epsilon_definition} in terms of the Buchdahl parameter $k$, which is an \emph{integration constant} of the $R^2$ vacuum field equations~\cite{Nguyen2022, Nguyen2023}, not a coupling constant of the theory.
Different central objects can, in principle, have different $k$ (and hence different $\epsilon$) values.
Moreover, $\tilde{k}=k/r_s$ depends on the Schwarzschild radius $r_s$, so even if $k$ were universal, $\epsilon$ would differ between the Sun and Sgr~A* due to their vastly different masses.

Second, the standard PPN formalism is not directly applicable to the Buchdahl-inspired metric in the usual sense.
In metric $f(R)$ gravity, when a well-defined Newtonian limit exists, $\gamma_{\rm PPN}$ is constrained to discrete values (typically 1 or 1/2, depending on the scalar-field mass and boundary conditions)~\cite{Toniato:2021dwi}, rather than admitting a continuous range as in scalar-tensor theories.
The asymptotically flat Buchdahl-inspired solution considered here is obtained as the $\Lambda\to 0$ limit of a more general family of metrics~\cite{Nguyen2022, Nguyen2023}. In this limit the Ricci scalar vanishes identically ($\mathcal{R}\equiv 0$), which differs from the generic metric $f(R)$ setting where a massive scalar degree of freedom mediates the post-Newtonian expansion. The relation $\gamma_{\rm PPN}=1-\epsilon$ should therefore be understood as an \emph{effective} mapping valid within the weak-field expansion of the specific Buchdahl solution, not as a universal PPN parameter to which solar-system bounds can be directly applied.

Consequently, the pulsar-timing forecast on $\epsilon$ around Sgr~A* probes a parameter space that is not already excluded by solar-system tests.
This is analogous to testing the no-hair theorem at Sgr~A*: the spacetime geometry near a specific black hole may deviate from the Kerr metric even if the theory reduces to GR in the solar system.

\subsection{Orbital dynamics}
\label{sec:orbital_dynamics}

We model the pulsar as a test particle moving in the gravitational field of the central black hole of gravitational mass $M_\bullet$ (i.e., the Newtonian gravitational mass parameter $\mu_{\rm N}$ in the notation of Sec.~\ref{sec:buchdahl_metric}).
The orbit is specified by the set of Keplerian parameters
\begin{equation}
    \bm{\eta}
    =
    \left\{
    P_b,\,
    e,\,
    i,\,
    \omega,\,
    \Omega,\,
    \theta_0
    \right\},
    \label{eq:orbital_parameters}
\end{equation}
where $P_b$ is the Keplerian period parameter (used to set the initial conditions via Kepler's third law; see below), $e$ is the initial eccentricity, $i$ is the inclination angle, $\omega$ is the initial argument of periastron, $\Omega$ is the longitude of the ascending node, and $\theta_0$ is the initial true anomaly. Without loss of generality, we set $\Omega=0$, since its value does not affect the TOAs in the spherically symmetric spacetime considered here~\cite{Taylor:1994zz}. (In the S2-like benchmark of Sec.~\ref{sec:S2like}, we list $\Omega=227^\circ$ for consistency with the S2 orbital elements, but the timing results are independent of this value.)

Given $P_b$ and $M_\bullet$, the semi-major axis $a$ is fixed by Kepler's third law,
\begin{equation}
    a^3 = \frac{M_\bullet\, P_b^2}{4\pi^2},
    \label{eq:kepler_third}
\end{equation}
which is used together with the remaining orbital elements $(e,\,i,\,\omega,\,\theta_0)$ to construct the initial position and velocity of the pulsar. Specifically, at the initial true anomaly $\theta_0$, the orbital radius is $r_0 = a(1-e^2)/(1+e\cos\theta_0)$, and the velocity is decomposed into radial and transverse components via the specific angular momentum $h = \sqrt{M_\bullet\, a(1-e^2)}$. These Newtonian initial conditions are then evolved under the 1PN equations of motion described below.

Following the weak-field expansion of the Buchdahl-inspired $R^2$ metric in Ref.~\cite{Yan2024}, the equation of motion is retained through first post-Newtonian (1PN) order,
\begin{equation}
    \frac{d^2\mathbf{r}}{dt^2}
    =
    -\frac{M_\bullet}{r^2}\mathbf{n}
    +\mathbf{a}_{\rm 1PN}^{R^2},
    \label{eq:eom_R2}
\end{equation}
where
\begin{equation}
    \mathbf{n}=\frac{\mathbf{r}}{r},
    \quad
    r=|\mathbf{r}|,
    \quad
    v^2=\mathbf{v}\cdot\mathbf{v},
    \quad
    \mathbf{v}=\frac{d\mathbf{r}}{dt}.
\end{equation}
The 1PN acceleration in the Buchdahl-inspired $R^2$ spacetime is
\begin{equation}
\begin{aligned}
    \mathbf{a}_{\rm 1PN}^{R^2}
    ={}&
    (4-2\epsilon)
    \frac{M_\bullet^2}{r^3}\mathbf{n}
    +(\epsilon-1)
    \frac{M_\bullet v^2}{r^2}\mathbf{n}
    \\
    &+
    (4-2\epsilon)
    \frac{M_\bullet}{r^3}
    \left(\mathbf{r}\cdot\mathbf{v}\right)\mathbf{v},
\end{aligned}
    \label{eq:a_1pn_R2}
\end{equation}
with $\epsilon$ the deviation parameter in Eq.~\eqref{eq:epsilon_definition}.
For clarity, Eq.~\eqref{eq:eom_R2} can be equivalently decomposed into a GR 1PN acceleration plus an $\epsilon$-dependent $R^2$ correction:
\begin{equation}
    \frac{d^2\mathbf{r}}{dt^2}
    =
    -\frac{M_\bullet}{r^2}\mathbf{n}
    +\mathbf{a}_{\rm 1PN}^{\rm GR}
    +\epsilon\,\mathbf{a}_{R^2},
    \label{eq:eom_decomposed}
\end{equation}
where
\begin{align}
    \mathbf{a}_{\rm 1PN}^{\rm GR}
    &=
    \frac{M_\bullet}{r^2}
    \left[
        \left(
            \frac{4M_\bullet}{r}-v^2
        \right)\mathbf{n}
        +4(\mathbf{n}\cdot\mathbf{v})\mathbf{v}
    \right],
    \label{eq:a_1pn_GR}\\
    \mathbf{a}_{R^2}
    &=
    -2\frac{M_\bullet^2}{r^3}\mathbf{n}
    +\frac{M_\bullet v^2}{r^2}\mathbf{n}
    -2\frac{M_\bullet}{r^3}
    \left(\mathbf{r}\cdot\mathbf{v}\right)\mathbf{v}.
    \label{eq:a_R2}
\end{align}
For $\epsilon=0$, the $R^2$ correction vanishes and the equations of motion reduce to the standard Schwarzschild 1PN dynamics.

\subsection{Timing model}
\label{sec:timing_model}

Our numerical timing scheme is based on the inverse-timing algorithm of Ref.~\cite{Hu:2023ubk} for a pulsar orbiting Sgr~A*.
Their model is formulated in GR using post-Newtonian orbital dynamics.
Here we retain the same numerical framework, but replace the GR orbital acceleration with the Buchdahl-inspired $R^2$ acceleration from Eq.~\eqref{eq:eom_R2}.
In addition, we modify the Shapiro delay according to the weak-field $R^2$ metric.

The pulsar rotational phase is described by
\begin{equation}
    N(T)
    =
    N_0+\nu T+\frac{1}{2}\dot{\nu}T^2,
    \label{eq:pulse_phase}
\end{equation}
where $T$ is the pulsar proper time, $N_0$ is an arbitrary phase offset, and $\nu$ and $\dot{\nu}$ are the pulsar spin frequency and its first time derivative.

The arrival time at a distant observer is related to the pulsar proper time by
\begin{equation}
    t_{\rm TOA}
    =
    T+\Delta_{\rm R}+\Delta_{\rm E}+\Delta_{\rm S},
    \label{eq:toa_model}
\end{equation}
where $\Delta_{\rm R}$, $\Delta_{\rm E}$, and $\Delta_{\rm S}$ denote the R\"omer, Einstein, and Shapiro delays, respectively.

Let $\hat{\mathbf K}_0$ be the unit vector pointing from the observer to Sgr~A*.
The R\"omer delay is
\begin{equation}
    \Delta_{\rm R}
    =
    \hat{\mathbf K}_0\cdot\mathbf{r}.
    \label{eq:roemer_delay}
\end{equation}

The Einstein delay describes the difference between coordinate time $t$ and pulsar proper time $T$,
\begin{equation}
    \Delta_{\rm E}\equiv t-T.
    \label{eq:einstein_delay}
\end{equation}
At 1PN order, we adopt the prescription
\begin{equation}
    \frac{dT}{dt}
    =
    \frac{
        1-\dfrac{M_\bullet}{r}-\dfrac{v^2}{2}
    }{
        1-\left\langle
        \dfrac{M_\bullet}{r}+\dfrac{v^2}{2}
        \right\rangle
    },
    \label{eq:proper_time_rate}
\end{equation}
where $v^2=\mathbf v\cdot\mathbf v$ and $\langle\cdots\rangle$ denotes an orbital average.
The denominator removes the secular component of $\Delta_{\rm E}$, which is degenerate with a redefinition of the pulsar spin frequency.
At the retained order, $\epsilon$ does not appear explicitly in Eq.~\eqref{eq:proper_time_rate}; it affects $\Delta_{\rm E}$ implicitly through the modified trajectory $\mathbf r(t)$ and velocity $\mathbf v(t)$.

The Shapiro delay is the propagation delay caused by the gravitational field of the central black hole.
In the GR timing model of Ref.~\cite{Hu:2023ubk}, the leading-order expression is
\begin{equation}
    \Delta_{\rm S}^{\rm GR}
    =
    -2M_\bullet
    \ln\left(
        \frac{r-\mathbf r\cdot\hat{\mathbf K}_0}{r_{\rm ref}}
    \right),
    \label{eq:shapiro_delay_GR}
\end{equation}
where $r_{\rm ref}$ is an arbitrary reference length; its choice only adds a constant phase offset and is therefore unobservable.

As shown in Eq.~\eqref{eq:shapiro_coefficient}, the null-geodesic analysis of the weak-field metric yields a Shapiro coefficient of $2-\epsilon$, arising from the temporal metric component $f(r)$ and the spatial metric component $g(r)$ respectively.
Replacing the GR coefficient $2$ in Eq.~\eqref{eq:shapiro_delay_GR} by $(2-\epsilon)$ gives
\begin{equation}
    \Delta_{\rm S}
    =
    -(2-\epsilon)M_\bullet
    \ln\left(
        \frac{r-\mathbf r\cdot\hat{\mathbf K}_0}{r_{\rm ref}}
    \right).
    \label{eq:shapiro_delay_R2}
\end{equation}
Equation~\eqref{eq:shapiro_delay_R2} reduces to the GR result when $\epsilon=0$.

For parameter estimation, it is computationally advantageous to calculate the pulse phase directly at specified arrival times.
Following Ref.~\cite{Hu:2023ubk}, we therefore use $t_{\rm TOA}$ as the independent integration variable.
Combining Eqs.~\eqref{eq:toa_model} and \eqref{eq:einstein_delay} gives
\begin{equation}
    t_{\rm TOA}
    =
    t+\Delta_{\rm R}(t)+\Delta_{\rm S}(t).
    \label{eq:toa_coordinate_time_relation}
\end{equation}
Taking the derivative yields
\begin{equation}
\begin{aligned}
    \frac{dt_{\rm TOA}}{dt}
    ={}&
    1+\hat{\mathbf K}_0\cdot\mathbf v
    \\
    &-(2-\epsilon)M_\bullet
    \frac{
        \mathbf n\cdot\mathbf v
        -\hat{\mathbf K}_0\cdot\mathbf v
    }{
        r-\mathbf r\cdot\hat{\mathbf K}_0
    },
\end{aligned}
    \label{eq:dttoa_dt_R2}
\end{equation}
where $\mathbf n=\mathbf r/r$.
Relative to the GR expression in Ref.~\cite{Hu:2023ubk}, the only explicit modification is the replacement $2\rightarrow 2-\epsilon$ in the Shapiro contribution.

The inverse timing equations are then
\begin{align}
    \frac{d\mathbf r}{dt_{\rm TOA}}
    &=
    \mathbf v\frac{dt}{dt_{\rm TOA}},
    \label{eq:inverse_orbit_equations1}\\
    \frac{d\mathbf v}{dt_{\rm TOA}}
    &=
     \ddot{\mathbf r}\frac{dt}{dt_{\rm TOA}},
    \label{eq:inverse_orbit_equations2}
\end{align}
with
\begin{equation}
    \frac{dt}{dt_{\rm TOA}}
    =
    \left(
        \frac{dt_{\rm TOA}}{dt}
    \right)^{-1},
    \label{eq:inverse_time_transformation}
\end{equation}
and
\begin{equation}
    \frac{d\Delta_{\rm E}}{dt_{\rm TOA}}
    =
    \left(
        1-\frac{dT}{dt}
    \right)
    \frac{dt}{dt_{\rm TOA}}.
    \label{eq:inverse_einstein_delay}
\end{equation}

After numerically integrating Eqs.~\eqref{eq:inverse_orbit_equations1}--\eqref{eq:inverse_einstein_delay}, the proper time associated with each arrival time is obtained from
\begin{equation}
    T
    =
    t_{\rm TOA}
    -\Delta_{\rm R}
    -\Delta_{\rm E}
    -\Delta_{\rm S}.
    \label{eq:proper_time_from_toa}
\end{equation}
Substituting $T$ into Eq.~\eqref{eq:pulse_phase} gives the predicted pulse phase $N(t_{\rm TOA})$ for a given set of timing-model parameters.

\section{Fisher Matrix Formalism}
\label{sec:fisher}

For a single pulsar, we denote the parameter vector by
\begin{equation}
    \bm{\lambda}
    =
    \left\{
    M_\bullet,\,
    \epsilon,\,
    P_b,\,
    e,\,
    i,\,
    \omega,\,
    \theta_0,\,
    N_0,\,
    \nu,\,
    \dot{\nu}
    \right\},
    \label{eq:lambda_vector}
\end{equation}
where $\epsilon$ is the Buchdahl-inspired $R^2$ deviation parameter.
The set $(N_0,\nu,\dot{\nu})$ contains nuisance parameters describing the intrinsic pulsar phase evolution and is included explicitly so that the Fisher forecasts account for correlations between $\epsilon$ and the spin-phase model. Here, $N_0$ is an arbitrary phase offset. Since it only fixes the origin of the pulse phase, we set $N_0=0$ without loss of generality. Because the intrinsic spin-frequency derivative is expected to be very small over the observational timescale considered here, we adopt $\dot{\nu}=0$ for the fiducial signal to simplify the calculation, while retaining $\dot{\nu}$ as a nuisance parameter in the Fisher analysis.

The numerical pipeline predicts the pulse phase $N(t_{\rm TOA})$ at a set of arrival times $\{t_k\}$.
We therefore construct the Fisher matrix in terms of phase measurements.
Assuming Gaussian timing noise corresponding to a TOA uncertainty $\sigma_{\rm TOA}$, the induced phase error for small timing perturbations satisfies $\delta N \simeq \nu\,\delta t$, so we take
\begin{equation}
    \sigma_N = \nu\,\sigma_{\rm TOA},
\end{equation}
evaluated at the fiducial spin frequency.

Let $N_k(\bm{\lambda})$ be the model-predicted pulse phase at the $k$-th TOA.
Then the Fisher information matrix is
\begin{equation}
    F_{ij}
    =
    \sum_{k=1}^{N_{\rm TOA}}
    \frac{1}{\sigma_N^2}\,
    \frac{\partial N_k}{\partial \lambda_i}\,
    \frac{\partial N_k}{\partial \lambda_j},
    \label{eq:fisher_def_phase}
\end{equation}
where derivatives are computed numerically (finite differences) or analytically depending on the implementation.

Under the standard Fisher approximation, the parameter covariance matrix is
\begin{equation}
    \mathbf{C} = \mathbf{F}^{-1},
\end{equation}
and the marginalized $1\sigma$ uncertainty for parameter $\lambda_i$ is
\begin{equation}
    \sigma_i = \sqrt{C_{ii}}.
\end{equation}
The correlation coefficient between parameters $\lambda_i$ and $\lambda_j$ is
\begin{equation}
    \rho_{ij} =
    \frac{C_{ij}}{\sqrt{C_{ii}C_{jj}}}.
\end{equation}

\section{Numerical Setup}
\label{sec:numerical_setup}

Our goal is to identify which classes of future pulsars are expected to yield the strongest sensitivity to the Buchdahl-inspired $R^2$ deviation parameter $\epsilon$.
To this end, we do not attempt a fit to any particular observed system. Instead, we carry out Fisher-matrix forecasts by systematically varying orbital parameters around a fiducial model.

The fiducial values are summarized in Table~\ref{tab:fiducial}. The baseline sampling interval is $\Delta t_{\rm samp}^{(0)}=7~{\rm d}$.

\begin{table}[h]
\centering
\caption{Fiducial parameters used in the Fisher forecast.}
\label{tab:fiducial}
\resizebox{\columnwidth}{!}{%
\begin{tabular}{l c c}
\toprule
\textbf{Parameter (unit)} & \textbf{Symbol} & \textbf{Fiducial value} \\
\midrule
Gravitational mass ($M_{\odot}$) & $M_\bullet$ & $4.3 \times 10^6$ \\
Keplerian period parameter (yr) & $P_b$ & $5$ \\
Eccentricity (initial) (dimensionless) & $e$ & $0.8$ \\
Inclination ($^\circ$) & $i$ & $60$ \\
Argument of periastron ($^\circ$) & $\omega$ & $90$ \\
Initial true anomaly ($^\circ$) & $\theta_0$ & $77$ \\
Spin frequency (Hz) & $\nu$ & $2$ \\
Spin-down rate ($\rm Hz\cdot s^{-1}$) & $\dot{\nu}$ & $0$ \\
Deviation parameter (dimensionless) & $\epsilon$ & $0$ \\
TOA uncertainty (s) & $\sigma_{\rm TOA}$ & $1\times 10^{-3}$ \\
\bottomrule
\end{tabular}%
}
\end{table}

\begin{figure}[h]
  \centering
  \includegraphics[width=\columnwidth]{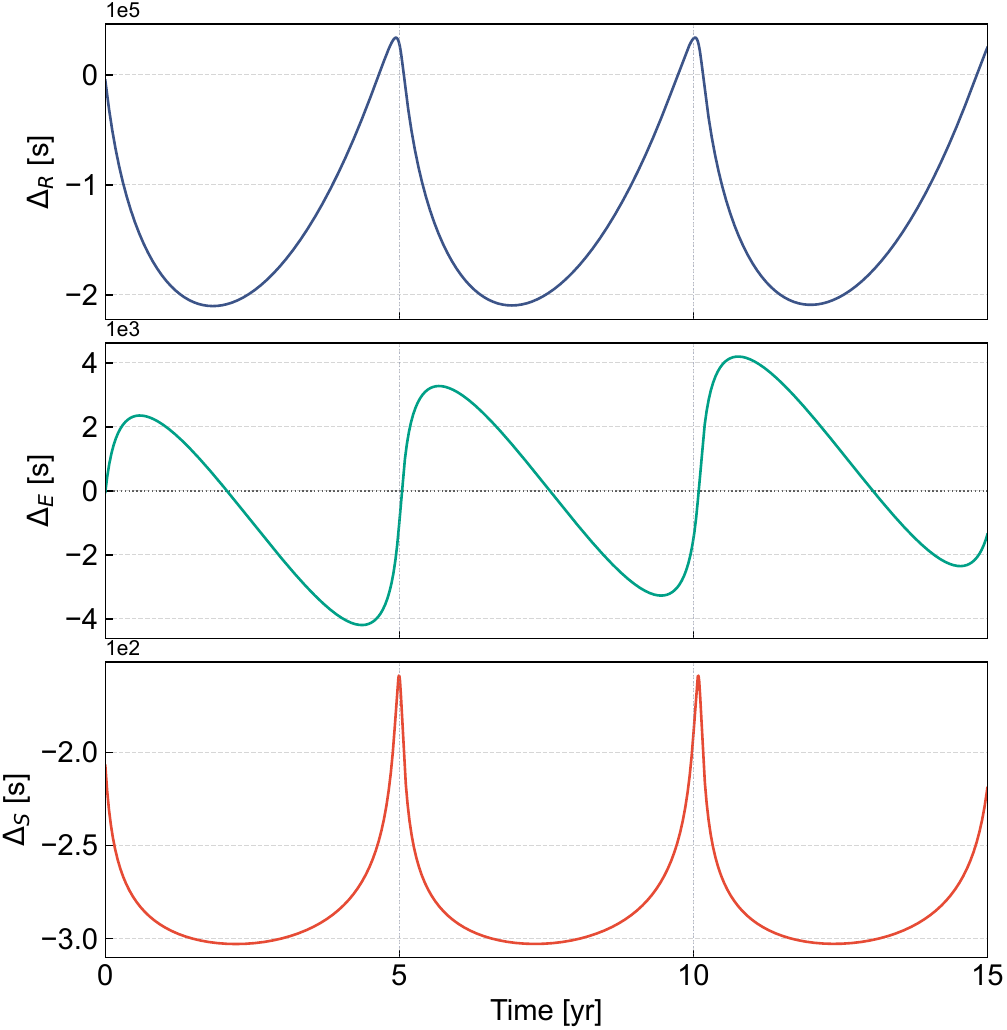}
  \caption{
  Fiducial timing delays for a pulsar orbiting Sgr~A* over $T_{\rm obs}=15~\mathrm{yr}$.
  The blue, green, and red curves show the R\"omer ($\Delta_{\rm R}$), Einstein ($\Delta_{\rm E}$), and Shapiro ($\Delta_{\rm S}$) delays, respectively.
  The vertical axes use units $\Delta_{\rm R}$ in $[10^{5}\,\mathrm{s}]$,
  $\Delta_{\rm E}$ in $[10^{3}\,\mathrm{s}]$,
  and $\Delta_{\rm S}$ in $[10^{2}\,\mathrm{s}]$.
  The secular trend in $\Delta_{\rm E}$ is removed in the timing prescription to avoid its near-degeneracy with the intrinsic spin-frequency evolution.
  }
  \label{fig:time_delay}
\end{figure}

\begin{figure}[h]
  \centering
  \includegraphics[width=\columnwidth]{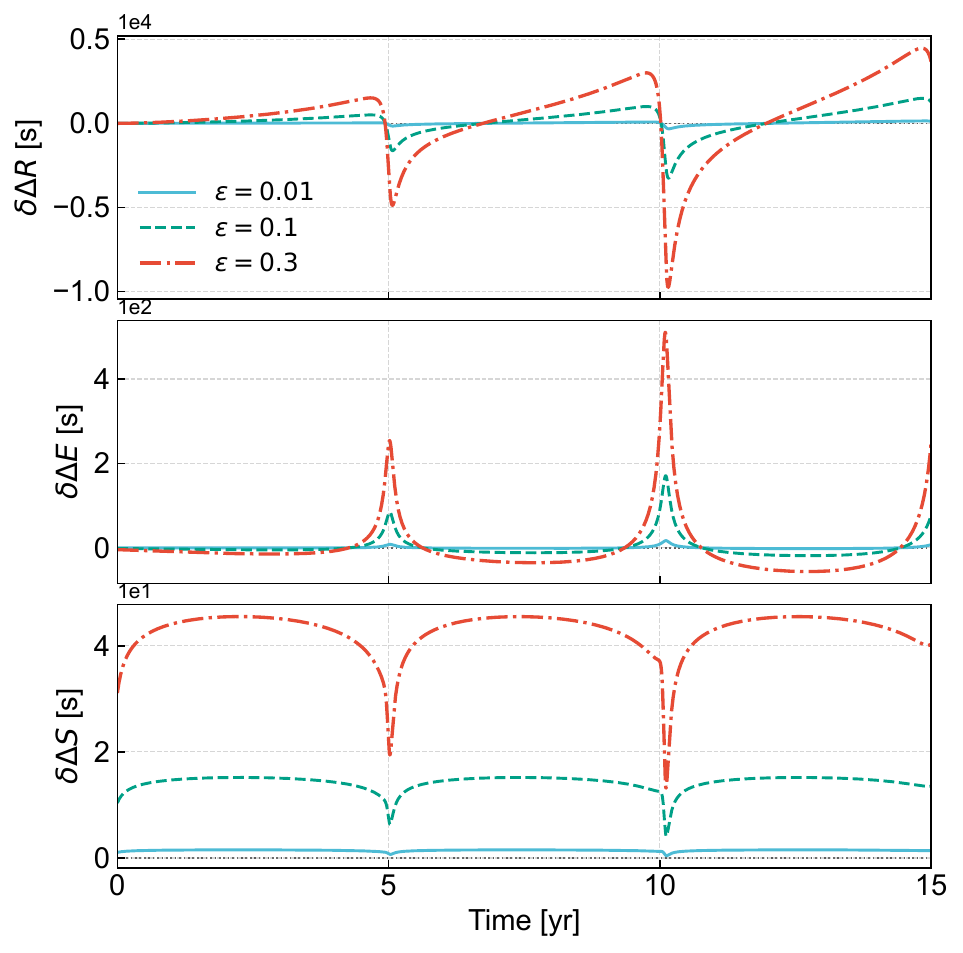}
  \caption{
  Differences between $R^2$-gravity and the GR fiducial case,
  $\delta\Delta\equiv \Delta(\epsilon)-\Delta(\epsilon=0)$, for $\epsilon=0.01,\,0.1,\,0.3$.
  Panels correspond to $\delta\Delta_{\rm R}$, $\delta\Delta_{\rm E}$, and $\delta\Delta_{\rm S}$.
  The vertical axes use units $\delta\Delta_{\rm R}$ in $[10^{4}\,\mathrm{s}]$,
  $\delta\Delta_{\rm E}$ in $[10^{2}\,\mathrm{s}]$,
  and $\delta\Delta_{\rm S}$ in $[10^{1}\,\mathrm{s}]$.
  The horizontal axis is time in $[\,\mathrm{yr}\,]$.
  }
  \label{fig:delta_delay}
\end{figure}

Since $\epsilon$ parametrizes deviations from GR, we use $\epsilon=0$ as the fiducial baseline.
For an observing time of $T_{\rm obs}=15~\mathrm{yr}$, the three timing delays are shown in Fig.~\ref{fig:time_delay}.
The blue, green, and red curves correspond to the R\"omer ($\Delta_{\rm R}$), Einstein ($\Delta_{\rm E}$), and Shapiro ($\Delta_{\rm S}$) delays, respectively.
To illustrate the impact of nonzero $\epsilon$, we define
\[
\delta\Delta \equiv \Delta(\epsilon) - \Delta(\epsilon=0),
\]
and plot $\delta\Delta$ for $\epsilon = 0.01,\,0.1,\,0.3$ in Fig.~\ref{fig:delta_delay}.

\section{Results}
\label{sec:results}
The Fisher-matrix forecasts presented below are designed to isolate how the marginalized uncertainty $\sigma_\epsilon$
depends on specific orbital factors. In Sec.~\ref{sec:precision_Pb} we vary the Keplerian period parameter, while
in Sec.~\ref{sec:precision_e} the eccentricity. These choices are motivated by the fact that in a highly eccentric
orbit the $\epsilon$-dependent timing signatures (from both the orbital dynamics and the Shapiro-delay coefficient)
are concentrated near periastron: changing $P_b$ and $e$ directly modifies the {\em cadence} and {\em strength} of the
near-periastron passages within the fixed observing window, thereby changing the information content for $\epsilon$
most efficiently. By contrast, variations in the line-of-sight projection of the orbit (e.g.\ the inclination $i$
and the argument of periastron $\omega$) mainly rephase the mapping between orbital motion and the observer’s
view, which tends to manifest primarily through parameter correlations and partial degeneracies (after
marginalization) rather than producing as direct an increase in the $\epsilon$-sensitive signal amplitude.
Section~\ref{sec:S2like} then assembles these trends into a benchmark forecast for a hypothetical pulsar on an
S2-like orbit, enabling a direct comparison with current stellar-orbit constraints.

\subsection{Keplerian period parameter}
\label{sec:precision_Pb}

The Keplerian period parameter $P_b$ controls both the size and temporal scale of the pulsar trajectory.
In the relativistic timing problem around Sgr~A*, this sets the effective accumulation of gravitational-field effects over the observing window, thereby influencing the Fisher information and the constraint on $\epsilon$.

A second consideration is that, for sufficiently tight (short-period) orbits, contributions associated with black-hole spin (frame dragging) can become increasingly important.
However, the Buchdahl-inspired metric adopted here is a weak-field, approximately isotropic, spherically symmetric modification and does not include gravitomagnetic effects.
To remain within the regime of validity of the metric approximation, we restrict the period parameter to
\begin{equation}
3~{\rm yr} \le P_b \le 10~{\rm yr}.
\end{equation}

\begin{figure}[h]
  \centering
  \includegraphics[width=\columnwidth]{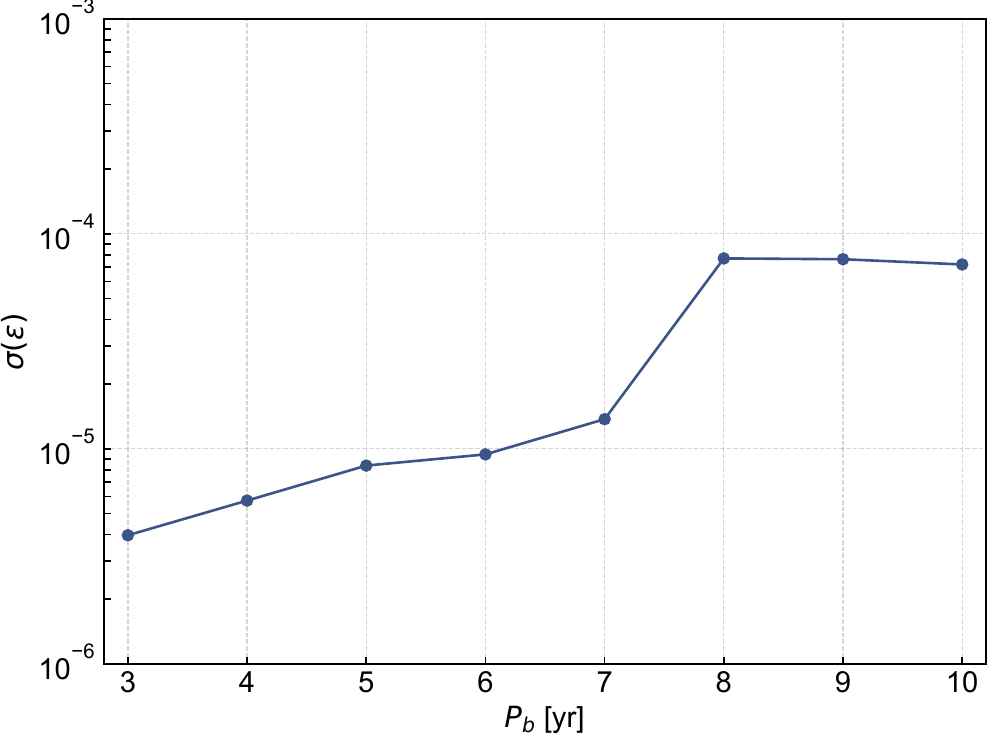}
  \caption{
  Forecasted $1\sigma$ precision on $\epsilon$ as a function of the Keplerian period parameter $P_b$ for $P_b\in[3,10]~\mathrm{yr}$.
  The curve is obtained from Fisher-matrix forecasts under the fiducial setup, with $T_{\rm obs}=15~\mathrm{yr}$.
  }
  \label{fig:precision_Pb}
\end{figure}

To avoid spurious effects from observing only a fraction of an orbit, we set $T_{\rm obs}=15~\mathrm{yr}$, so that the data set typically covers multiple orbital phases across the considered $P_b$ range.
We then compute the resulting precision on $\epsilon$ for $P_b\in[3,10]~\mathrm{yr}$.
In these scans, the scanned parameter (e.g., $P_b$) is varied while the other parameters are held at their fiducial values; the quoted uncertainty is obtained by inverting the full Fisher matrix (including all parameters) at each scan point, so that correlations with all other parameters are accounted for.
As shown in Fig.~\ref{fig:precision_Pb}, the forecasted uncertainty on $\epsilon$ improves systematically for shorter orbital periods. This trend reflects the combined effect of several factors that are not separately disentangled in the present analysis: shorter $P_b$ implies a smaller semi-major axis (and hence stronger gravitational fields near periastron), a larger number of orbital cycles within the fixed observing window, and more frequent periastron passages. A controlled decomposition of these contributions would require a phase-resolved Fisher information analysis, which is beyond the scope of this work.

\subsection{Eccentricity}
\label{sec:precision_e}

We next investigate how the fiducial orbital eccentricity affects the
constraint on $\epsilon$. For fixed period parameter $P_b$ and fixed system
masses, the semi-major axis remains approximately fixed, while the
periastron and apastron distances vary as
\begin{equation}
    r_{\rm p}=a(1-e),
    \qquad
    r_{\rm a}=a(1+e).
\end{equation}
Increasing $e$ therefore brings the pulsar closer to the black hole near
periastron, where relativistic corrections to both the orbital dynamics
and photon propagation are generally enhanced.

This observation suggests that highly eccentric systems may carry more
information about $\epsilon$. However, the resulting improvement is not
guaranteed a priori. The pulsar spends a smaller fraction of each orbit
near periastron as $e$ increases, and the final constraint also depends on
the observing cadence and on correlations between $\epsilon$ and the
remaining model parameters.

To illustrate the phase dependence, the data may be divided into
periastron- and apastron-dominated subsets. For statistically independent
timing measurements, their Fisher matrices satisfy
\begin{equation}
    \mathbf{F}=\mathbf{F}_{\rm p}+\mathbf{F}_{\rm a}.
\end{equation}
Nevertheless, the uncertainty
\begin{equation}
    \sigma_\epsilon
    =\sqrt{\left(\mathbf{F}^{-1}\right)_{\epsilon\epsilon}},
\end{equation}
cannot in general be decomposed into independent periastron and apastron
contributions, because it depends on the full covariance structure.
The competition between enhanced periastron sensitivity, reduced
residence time, and parameter degeneracies must therefore be evaluated
using the complete Fisher matrix.

\begin{figure}[h]
  \centering
  \includegraphics[width=\columnwidth]{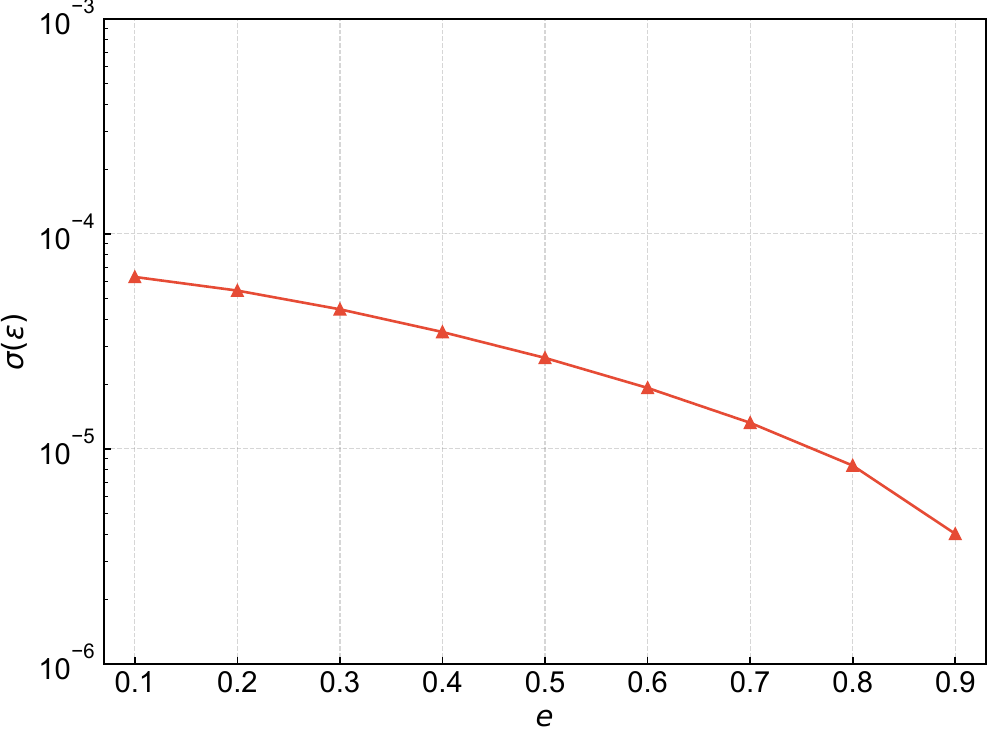}
  \caption{$1\sigma$ uncertainty on $\epsilon$ as a function of the fiducial orbital eccentricity $e$, with $P_b$, the system masses, and the observational setup held at their fiducial values. All parameters participate in the marginalization. The decrease in $\sigma_\epsilon$ is consistent with enhanced near-periastron sensitivity dominating over the reduced residence time and changes in parameter correlations.
}
  \label{fig:precision_e}
\end{figure}

We perform Fisher-matrix forecasts for fiducial eccentricities in the
range $e\in[0.1,0.9]$, while keeping $P_b$, the system masses, and the
observational setup fixed. As shown in Fig.~\ref{fig:precision_e}, the
$1\sigma$ uncertainty on $\epsilon$ decreases monotonically
over the range considered. For the adopted cadence and timing model, the
enhancement of the $\epsilon$-dependent signatures near periastron
therefore dominates over the reduced time spent in that orbital phase
and any changes in parameter correlations.

\subsection{Benchmark: S2-like pulsar}
\label{sec:S2like}

Having established how individual factors influence the forecast precision, we now assemble these insights into a benchmark forecast designed to enable a direct comparison with current stellar-orbit constraints.
The S2 star is the best-studied object orbiting Sgr~A*, and our previous work~\cite{Yan2024} used its astrometric and spectroscopic data to constrain $\epsilon$.
Here, we ask: \emph{if a pulsar were found on an S2-like orbit, how much better could it constrain $\epsilon$?}

We consider a hypothetical pulsar moving around Sgr~A* on an S2-like orbit with parameters approximately reproducing those of the S2 star~\cite{GRAVITY2018,GRAVITY2,Mon},
\begin{equation}
\begin{gathered}
    P_b = 16~{\rm yr}, \qquad
    e = 0.88, \qquad
    i = 135^\circ, \\
    \omega = 65^\circ, \qquad
    \Omega = 227^\circ, \qquad
    \theta_0 = 0.
\end{gathered}
\end{equation}
Here $\theta_0=0$ means the observing window starts at periastron.
The pulsar-specific parameters and observational assumptions are set to the fiducial values listed in Table~\ref{tab:fiducial}. We take $\epsilon=0$ as the fiducial signal and adopt a uniform timing cadence of seven days.

\begin{figure}[h]
    \centering
    \includegraphics[width=\columnwidth]
        {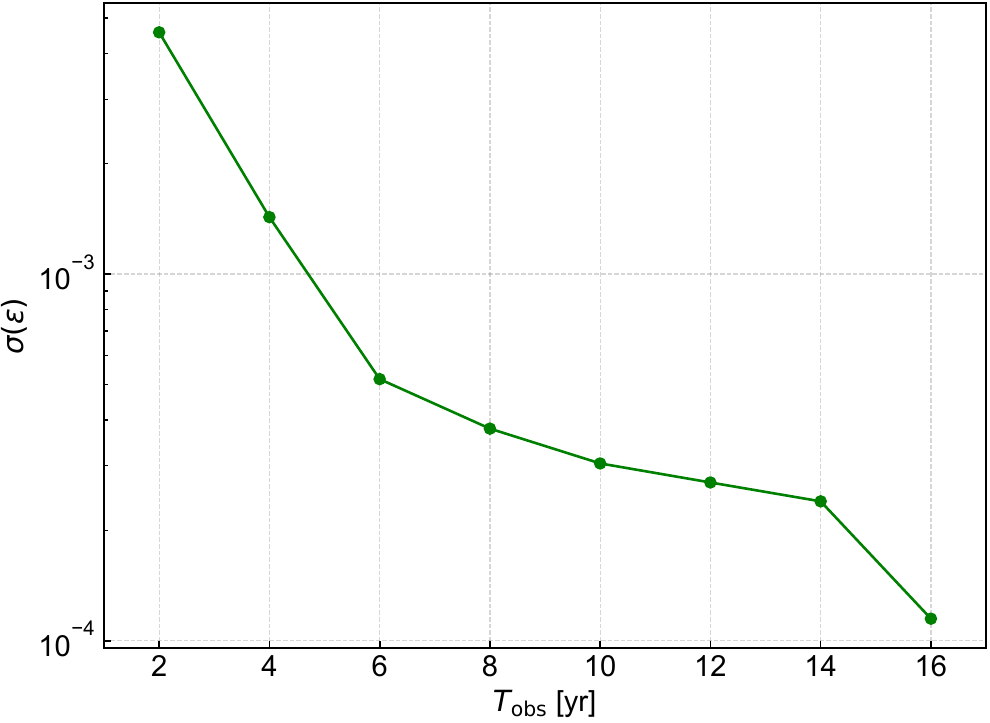}
    \caption{
        $1\sigma$ uncertainty on $\epsilon$ as a function of the observing span for the fiducial S2-like pulsar, obtained from the Fisher matrix.
        The timing cadence is fixed at seven days.
    }
    \label{fig:S2like_Tobs}
\end{figure}

Using the Fisher-matrix procedure described in Sec.~\ref{sec:fisher}, we calculate the parameter uncertainties for observing spans between $T_{\rm obs}=2~{\rm yr}$ and $16~{\rm yr}$.
For each observing span, we marginalize over the black-hole mass, orbital parameters, reference pulse phase, spin frequency, and spin-frequency derivative.
The resulting dependence of $\sigma_\epsilon$ on the observing span is shown in Fig.~\ref{fig:S2like_Tobs}.

The constraint generally improves as the observing baseline increases.
This improvement arises both from the increasing number of pulse times of arrival (TOAs) and from the broader coverage of the orbital phase.
In particular, the high eccentricity of the orbit causes the orbital and pulse-propagation signatures to vary rapidly near periastron, where a substantial fraction of the timing information is expected to accumulate, although a rigorous decomposition of Fisher information by orbital phase has not been performed here.
Observing a larger fraction of the orbit also helps distinguish the effect of $\epsilon$ from variations in the orbital and spin parameters.

For the fiducial S2-like orbit ($P_b=16~{\rm yr}$), we set $T_{\rm obs}=16~{\rm yr}$, noting that $P_b$ is the Keplerian period parameter used to set the initial conditions (Sec.~\ref{sec:orbital_dynamics}), not the fully relativistic radial period; the Fisher-matrix forecast yields a sensitivity of order
\begin{equation}
    \sigma_\epsilon \sim 10^{-4}.
\end{equation}
Table~\ref{tab:S2like_sigma} summarizes the marginalized $1\sigma$ uncertainties on all ten fitting parameters for this benchmark configuration.

\begin{table}[h]
\centering
\caption{Marginalized $1\sigma$ Fisher-matrix uncertainties for the S2-like pulsar benchmark ($P_b=16~{\rm yr}$, $e=0.88$, $T_{\rm obs}=16~{\rm yr}$, cadence $=7~{\rm d}$, $\sigma_{\rm TOA}=1~{\rm ms}$).}
\label{tab:S2like_sigma}
\resizebox{\columnwidth}{!}{%
\begin{tabular}{l c c c}
\toprule
\textbf{Parameter (unit)} & \textbf{Symbol} & \textbf{Fiducial value} & $\sigma$ ($1\sigma$) \\
\midrule
Gravitational mass ($M_{\odot}$) & $M_\bullet$ & $4.3\times10^{6}$ & $4.41$ \\
Keplerian period parameter (yr) & $P_b$ & $16.0$ & $2.92\times10^{-5}$ \\
Eccentricity (initial) (dimensionless) & $e$ & $0.88$ & $1.45\times10^{-7}$ \\
Inclination ($^\circ$) & $i$ & $135$ & $1.88\times10^{-5}$ \\
Argument of periastron ($^\circ$) & $\omega$ & $65$ & $7.62\times10^{-6}$ \\
Initial true anomaly ($^\circ$) & $\theta_0$ & $0$ & $6.59\times10^{-6}$ \\
Phase offset (dimensionless) & $N_0$ & $0$ & $1.98\times10^{-2}$ \\
Spin frequency (Hz) & $\nu$ & $2.0$ & $1.58\times10^{-10}$ \\
Spin-down rate ($\rm Hz \cdot s^{-1}$) & $\dot{\nu}$ & $0$ & $3.09\times10^{-19}$ \\
Deviation parameter (dimensionless) & $\epsilon$ & $0$ & $1.15\times10^{-4}$ \\
\bottomrule
\end{tabular}%
}
\end{table}

In our previous work~\cite{Yan2024}, we used the astrometric and radial-velocity data of the S2 star to constrain the Buchdahl-inspired $R^2$ deviation parameter, obtaining a 95\% confidence interval half-width of $|\epsilon|_{\rm S2}^{\rm 95\%}\approx 0.56$.
The pulsar-timing forecast $\sigma_\epsilon\sim 10^{-4}$ is a $1\sigma$ statistical sensitivity, and is therefore not directly comparable to the S2 95\% interval on an equal footing.
Nevertheless, under the adopted idealized assumptions, the characteristic statistical scale is several orders of magnitude below the current S2 interval, suggesting that pulsar timing has the potential to probe $\epsilon$ at a substantially deeper level than the stellar-orbit method.
This expectation is attributable to two factors: (i) the phase-coherent nature of pulsar timing, which effectively integrates timing information over the full observing span with millisecond-level TOA precision, and (ii) the explicit $\epsilon$-dependence of the Shapiro-delay coefficient [Eq.~\eqref{eq:shapiro_delay_R2}], which provides a photon-propagation signature not available as a phase-coherent observable in conventional stellar astrometry and spectroscopy.

Finally, this calculation is an idealized forecast for a hypothetical system.
It assumes Gaussian and uncorrelated TOA uncertainties and neglects red timing noise, pulse-profile evolution, interstellar scattering, dispersion-measure variations, and environmental perturbations near the Galactic center.
Moreover, no pulsar on an S2-like orbit around Sgr~A* is currently known.
The results should therefore be interpreted as prospective sensitivities under the adopted observational assumptions.

\subsection{Domain of validity of the weak-field forecast}
\label{sec:model_validity}

The Buchdahl-inspired metric employed in this work is used here only through its weak-field expansion at first post-Newtonian (1PN) order. Accordingly, we do not supplement the timing model with phenomenological 2PN terms, since our analysis is restricted to the 1PN-accurate expansion consistently derived within the Buchdahl-inspired framework. Likewise, the spacetime is assumed to be static and spherically symmetric, so black-hole spin and gravitomagnetic effects are outside the scope of the present model.

The reported Fisher uncertainties should consequently be interpreted as statistical sensitivities within the adopted weak-field, spherically symmetric timing framework, rather than as complete estimates of the accuracy achievable for the real Sgr~A* system. For the S2-like orbit, the largest weak-field expansion parameter occurs near periastron and is approximately
\begin{equation}
    u_{\rm p}
    \equiv
    \frac{M_\bullet}{r_{\rm p}c^2}
    \sim {\rm few}\times10^{-4}.
\end{equation}
The retained Buchdahl-inspired correction scales schematically as $\epsilon\, u_{\rm p}$ relative to the Newtonian dynamics, whereas the first omitted terms in a generic post-Newtonian expansion are expected to scale as $u_{\rm p}^2$. A simple order-of-magnitude comparison therefore gives
\begin{equation}
    \epsilon_{\rm trunc}
    \sim u_{\rm p}
    \sim \mathcal{O}(10^{-4}).
\end{equation}
This estimate is not a rigorous theoretical-error bound, because the coefficients and time dependence of the higher-order Buchdahl-inspired terms are presently unknown. Nevertheless, it shows that the sensitivity $\sigma_\epsilon\sim 10^{-4}$ approaches the natural truncation scale of the weak-field model.

Applying the forecast to actual observations of a pulsar around Sgr~A* would require either a higher-order extension of the Buchdahl-inspired timing model or an explicit treatment of theoretical uncertainty. It would also require the inclusion of black-hole spin and other environmental effects, which may be present in the real Galactic-center system even though they are absent by construction from the static metric considered here.

\section{Conclusions}
\label{sec:conclusion}

We performed Fisher-matrix forecasts for single-pulsar timing around Sgr~A* in a Buchdahl-inspired $R^2$ gravity model parameterized by $\epsilon$.
Using a timing framework that accounts for both orbital dynamics and light-propagation delays, we investigated how the constraint on $\epsilon$ depends on orbital parameters.
Our main findings are:

\begin{itemize}
    \item \textbf{Shorter orbital periods} generally improve sensitivity to $\epsilon$, reflecting stronger accumulation of relativistic timing signatures over the observing window.
    \item \textbf{Larger eccentricities} yield tighter constraints, consistent with enhanced sensitivity from observations near periastron, although a phase-resolved Fisher information decomposition has not been performed.
    \item \textbf{S2-like pulsar benchmark:} For an idealized pulsar on an orbit approximately resembling that of S2 ($P_b=16~\rm yr$, $e=0.88$), the Fisher-matrix forecast yields a sensitivity of order $\sigma_\epsilon\sim 10^{-4}$.
    Under the adopted idealized assumptions, the characteristic statistical scale is several orders of magnitude below the current S2 stellar-orbit 95\% confidence interval half-width ($|\epsilon|_{\rm S2}^{\rm 95\%}\approx 0.56$)~\cite{Yan2024}, though this comparison is heuristic given the differing confidence levels.
\end{itemize}

These trends provide guidance for target selection and observing planning for future Galactic-center pulsar programs, aiming to maximize the information content relevant to $R^2$-type deviations from GR.
In particular, the most favorable targets are short-period, high-eccentricity pulsars.

We emphasize that the Buchdahl-inspired timing model is presently formulated at 1PN order and does not include black-hole spin or other higher-order effects. The reported Fisher sensitivities should therefore be understood as statistical forecasts within this truncated weak-field, spherically symmetric framework. The sensitivity $\sigma_\epsilon\sim 10^{-4}$ approaches the natural truncation scale of the 1PN expansion ($\sim u_{\rm p}\sim 10^{-4}$), beyond which unmodeled higher-order terms would need to be included. Applying these forecasts to actual observations of a pulsar around Sgr~A* would require a higher-order extension of the Buchdahl-inspired timing model, as well as explicit treatment of black-hole spin, environmental perturbations, and other systematic effects.

\begin{acknowledgments}
In this work, Jian-Ming Yan is supported by the China Postdoctoral Science Foundation Grant No.~2026M793675. Tao Zhu is supported by the National Natural Science Foundation of China under Grants No. 12275238, No. 12675080,  No. 12542053, and No. 11675143, the National Key Research and Development Program under Grant No. 2020YFC2201503, and the Zhejiang Provincial Natural Science Foundation of China under Grants No. LR21A050001 and No. LY20A050002, and the Fundamental Research Funds for the Provincial Universities of Zhejiang in China under Grant No. RF-A2019015.

We thank Lijing Shao and Zexin Hu (Peking University) for helpful comments.
\end{acknowledgments}

\end{document}